\documentstyle[11pt,epsf]{article}

\def\eps{\epsilon}
\def\Ord{{\cal O}}
\def\oneloop{{1 \mbox{-} \rm loop}}
\def\tree{{\rm tree}}

\def\Gr{{\rm Gr}}
\def\Tr{{\rm Tr}}
\def\cg{c_\Gamma}
\def\spa#1.#2{\left\langle#1\,#2\right\rangle}
\def\spb#1.#2{\left[#1\,#2\right]}
\def\Split{\mathop{\rm Split}\nolimits}

\def\N{G^n}
\def\F{G^f}

\begin{document}
\topskip 1cm 
\begin{titlepage}

\hspace*{\fill}\parbox[t]{4cm}{
EDINBURGH 98/24 \\
MSUHEP-90324 \\ UCLA/99/TEP/12 \\ BNL-HET-99/9}

\vspace{.1cm}

\begin{center}
{\Large\bf The Infrared Behavior of QCD Cross Sections at
Next-to-Next-to-Leading Order} \\
\vspace{.8cm}

{Zvi Bern}\\
\vspace{.2cm}
{\sl Department of Physics and Astronomy\\
University of California at Los Angeles\\
Los Angeles,  CA 90095-1547, USA}\\

\vspace{.6cm}
{Vittorio Del Duca\footnote{On leave of absence from
I.N.F.N., Sezione di Torino, Italy.}\footnote{Rapporteur at the
Corfu Summer Institute on Elementary Particle Physics, 1998.}}\\
\vspace{.2cm}
{\sl Particle Physics Theory Group\\
Dept. of Physics and Astronomy\\ University of Edinburgh\\
Edinburgh EH9 3JZ, Scotland, UK}\\

\vspace{.6cm}
{William B. Kilgore}\\
\vspace{.2cm}
{\sl Department of Physics\\
Brookhaven National Laboratoy\\
Upton, NY 11973-5000, USA}\\

\vspace{.4cm}
and \\
\vspace{.4cm}

{Carl R. Schmidt}\\
\vspace{.2cm}
{\sl Department of Physics and Astronomy\\
Michigan State University\\
East Lansing, MI 48824, USA}\\

\vspace{.4cm}

\begin{abstract}
In this talk we examine how one-loop soft and collinear 
splitting functions occur in the calculation of 
next-to-next-to-leading order (NNLO) corrections to
production rates, and we present the one-loop gluon soft and 
splitting functions, computed to all orders in the
dimensional regularization parameter $\epsilon$. 
We apply the one-loop gluon soft function to the calculation
of the next-to-leading logarithmic corrections to the Lipatov vertex 
to all orders in $\epsilon$.
\end{abstract}
\end{center}
\vfil

\end{titlepage}

The single most important parameter of perturbative QCD is the strong 
coupling constant, $\alpha_s$, which has been determined in several
ways~\cite{schm}. Some of the most promising ones are due to hadron
production in $e^+ e^-$ collisions; e.g., the hadronic branching
ratio of the $Z^0$ or global event shape variables in $e^+ e^- \rightarrow
3$ jets. The hadronic branching ratio $R_Z$ is known in perturbative QCD
to three loops; however, the usefulness of this observable in the 
determination of $\alpha_s$ is
limited by the sensitivity of $R_Z$ to other Standard 
Model parameters~\cite{hag} (for an overview, see ref.~\cite{stef}).
On the contrary, $e^+ e^- \rightarrow 3$ jets, which is known only 
to next-to-leading order (NLO)~\cite{ert,nk},
does not suffer from the above limitations. Thus a 
next-to-next-to-leading order (NNLO) calculation
of this process could yield a significant
reduction of the theoretical uncertainty 
in the determination of $\alpha_s$.

In order to understand the general features of a calculation at NNLO, 
we begin by outlining how a higher-order calculation of a scattering process
is performed.  At leading order (LO) in $\alpha_s$ the cross section
is obtained by squaring the tree amplitudes.  If
$n$ particles are produced in the scattering,  
each of them will be resolved in the final state. 
Thus no singularities appear in the LO cross section.
At LO the coupling  $\alpha_s$ is evaluated with one-loop running, so that
there is an implicit  dependence on an arbitrary renormalization scale 
$\mu_R$.  In addition,
if one or both of the scattering particles are strongly interacting,
the cross section will factorize into the convolution of parton
density functions (to be determined experimentally)
and a hard partonic cross section, which is computed as an expansion 
in $\alpha_s$.  This procedure introduces into both the parton densities 
and the partonic cross section a dependence on a second arbitrary parameter,
the factorization scale $\mu_F$ \cite{css}. Typically, the dependence
on $\mu_R$ and $\mu_F$ is maximal at LO.

The calculation of the cross section at next-to-leading order (NLO) in 
$\alpha_s$ is less straightforward.
Two series of amplitudes are required in the squared matrix elements: 
$a)$ tree and one-loop amplitudes for the
production of $n$ particles; $b)$ tree amplitudes for the
production of $n+1$ particles. The one-loop amplitudes typically
have virtual ultraviolet and infrared singularities, which may be 
regularized using dimensional regularization.  This involves
analytically continuing the loop momenta into
$D=4-2\epsilon$ dimensions, so that the one-loop amplitude is now a 
function of $\epsilon$.
If this is expanded in $\epsilon$, the ultraviolet singularities
appear as single poles in $\epsilon$, which can be removed by
renormalizing the amplitude.  This introduces an explicit dependence on 
the renormalization scale $\mu_R$. 

At NLO the structure of the infrared singularities has been extensively
studied. Virtual infrared singularities appear as
double poles in $\epsilon$, when they are both soft and collinear,
and single poles in $\epsilon$, when they are either soft or
collinear. Real infrared singularities occur in the phase-space integral
over the $n+1$ final-state particles of the squared tree
amplitudes, either when any gluon becomes soft or when any two massless
particles become collinear, thus yielding single poles in $\epsilon$. If
one of the two collinear particles is soft, a double pole 
in $\epsilon$ arises. The singularities occur in a universal way, i.e.
independent of the particular amplitude
considered. Accordingly, soft singularities can be accounted for by
universal tree soft functions \cite{bcm,bg}, and collinear
singularities by universal tree splitting functions \cite{ap}.
These have also been combined into a single function~\cite{dipole}.
A detailed discussion of the
infrared singularities at NLO for $e^+ e^- \rightarrow$ jets may be
found, for example, in ref.~\cite{gg}.

For processes with no strongly interacting scattering particles,
all infrared divergences cancel when real and virtual contributions 
are put together to form the NLO coefficient in the expansion of the
cross section \cite{kln}.  Typically, 
the dependence on $\mu_R$ is reduced at NLO.
For processes with strongly-interacting scattering particles,
all infrared divergences cancel except for those associated with
initial-state collinear singularities, which manifest themselves
as single poles in $\epsilon$; these singularities are factorized 
into the parton densities, thus reducing the dependence of the
cross section on $\mu_F$ \cite{css}.

In order to compute a cross section at NNLO, three series of amplitudes are
required: {\it a}) tree, one-loop, and two-loop amplitudes
for the production of $n$ particles; {\it b}) 
tree and one-loop amplitudes for the
production of $n+1$ particles; {\it c}) tree amplitudes for the
production of $n+2$ particles.  For the case of NNLO $e^+ e^-
\rightarrow 3$ jets the five-parton final-state
tree~\cite{ZFiveJetsBorn} amplitudes, as well as the four-parton final-state 
one-loop amplitudes exist in both helicity~\cite{bdkZ4} and
squared matrix-element forms~\cite{cgmZ4}. However, as we discuss below, in
order to be used in NNLO computations higher-order terms in $\epsilon$
must be included. For the required two-loop three-parton final-state
amplitudes no computations exist, as yet. 
For single- and double-jet production at hadron
colliders
the six-parton tree~\cite{bg87,xu} amplitudes, as well as
the five-parton one-loop amplitudes~\cite{bdk5g,kst1g4q,bdk3g2q} exist
in helicity matrix-element form, but no four-parton two-loop amplitude
computations exist, as yet. Indeed, no two-loop amplitude
computations exist for cases containing more than a single
kinematic variable, except in the special cases of maximal
supersymmetry~\cite{byr}.

In the calculation of a production rate at NNLO the structure of the 
infrared singularities is the following:
\begin{description}
\item{\it i)\ } In the squared tree amplitudes, any
two of the $n+2$ final-state particles can be unresolved. Accordingly the
ensuing soft singularities, collinear singularities, and mixed
collinear/soft singularities have been accounted for by double-soft
functions \cite{bg}, double-splitting functions and soft-splitting
functions \cite{cg}, respectively. 
\item{\it ii)\ } In the interference term between a two-loop amplitude for 
the production of $n$ particles and its tree-level counterpart,
all the produced particles are resolved in the 
final state and no new singularities appear through the phase-space 
integration.  Thus,  the expansion of the two-loop amplitude in 
$\epsilon$, which starts with a $1/\epsilon^4$ pole, can be truncated at
$O(\epsilon^0)$.  The universal
structure of the coefficients of the $1/\eps^4$, $1/\eps^3$ and
$1/\eps^2$ poles has been determined \cite{cat}.
\item{\it iii)\ } In the interference term
between a one-loop amplitude for the production of $n+1$ particles and
its tree-level counterpart any one of the 
produced particles can be unresolved in the final state; hence, the
phase-space integration gives at most an additional double
pole in $\eps$.  Therefore, the expansion in $\epsilon$ of the
interference term starts with a $1/\epsilon^4$ pole, from mixed
virtual/real infrared singularities, and in order to evaluate it to
$\Ord(\epsilon^0)$, the $(n+1)$-parton one-loop amplitude needs to
be evaluated to $\Ord(\epsilon^2)$.  (A similar need to evaluate
one-loop amplitudes to higher orders in $\epsilon$ has been previously
noted in NNLO deep inelastic scattering \cite{van} and in the 
next-to-leading-logarithmic (NLL)
corrections to the BFKL equation \cite{ffk}.)  
\item{\it iv)\ } In the square of the one-loop amplitude for the
production of $n$ particles, the expansion in $\epsilon$ of the
amplitude, which starts with a
$1/\epsilon^2$ pole, must be known to $\Ord(\epsilon^2)$ in order to
evaluate the squared amplitude to $\Ord(\epsilon^0)$.
\end{description}

Here we focus on the singularities in {\it iii)}, 
which require that 
the $(n+1)$-parton one-loop amplitudes be evaluated to 
$\Ord(\epsilon^2)$.  
For the case of NNLO
corrections to $e^+ e^- \rightarrow 3$ jets and to single- and 
double-jet production at hadron colliders, this would be a rather
formidable task given the already non-trivial analytic structure of the
one-loop $e^+ e^- \rightarrow 4$ partons amplitudes~\cite{bdkZ4,cgmZ4} 
and of the one-loop five-parton amplitudes~\cite{bdk5g,kst1g4q,bdk3g2q},
both presented through $\Ord(\eps^0)$ only.
However, a simplification can be made if one uses the fact that
the additional double poles in $\eps$ of the interference term
arise from the infrared-divergent regions of the phase-space integration.
This implies that the one-loop $(n+1)$-parton final-state amplitude needs be 
calculated to $\Ord(\epsilon^2)$ only in the regions where two partons 
become collinear or one parton becomes soft.
Therefore, one can use this amplitude calculated to $\Ord(\epsilon^0)$
and then supplement it in the soft or collinear regions by appropriate
$\Ord(\epsilon^2)$ terms.
In these regions the amplitude factorizes into sums
of products of $n$-parton final-state amplitudes multiplied by soft or
collinear splitting functions. It is these soft or collinear
splitting functions and the one-loop $n$-parton final-state amplitudes
that must be evaluated to $\Ord(\epsilon^2)$. This is a much
simpler task than evaluating the full one-loop $(n+1)$-parton final-state
amplitudes beyond $\Ord(\eps^0)$.

Below, we provide the one-loop gluon splitting and soft
functions to all orders in $\epsilon$~\cite{bds}\footnote{
The one-loop splitting functions through
$\Ord(\epsilon^0)$ can be found in \cite{bddkSusy4,bdk3g2q}, 
and the one-loop soft
functions through $\Ord(\epsilon^0)$ may be extracted from the known
four- \cite{bk} and five-parton \cite{bdk5g,kst,bdk3g2q} one-loop
amplitudes.}.
A complete listing of the one-loop splitting and soft
functions, including fermions, is given elsewhere~\cite{bdks}.
Then we apply the framework outlined above
to one of the effective vertices of the NLL
corrections~\cite{ff} to the BFKL equation~\cite{bfkl}, 
namely to the one-loop amplitude for 
three-parton production in multi-Regge kinematics~\cite{fl,ffk,dds}, for which
the produced partons are strongly ordered in rapidity. In NNLO and in NLL
corrections to two-jet scattering,
this amplitude appears in an interference term multiplied by its 
tree-level counterpart. Because of the rapidity ordering
in the multi-Regge kinematics, the
phase-space integration does not yield any collinear singularities;
however, the gluon which is intermediate in rapidity can become soft.
Accordingly the one-loop amplitude must be determined to $O(\epsilon^0)$
plus the contribution with the soft intermediate gluon evaluated to
$O(\epsilon)$ \cite{ffk,dds}. To determine the soft
gluon contribution we use our all orders in $\eps$ determination of
the soft functions together with previous all orders in $\eps$
determinations of the four-gluon amplitudes~\cite{gsb,bddkSusy4,bm}.

We first briefly review properties of $n$-gluon scattering amplitudes.
The tree-level color decomposition is (see
e.g.\ ref.\cite{mpReview} for details and normalizations)
\begin{equation}
M_n^\tree(1,2,\ldots n) =  g^{(n-2)} \sum_{\sigma \in S_n/Z_n}
{\rm Tr}\left( T^{a_{\sigma(1)}} 
T^{a_{\sigma(2)} }\cdots T^{a_{\sigma(n)}} \right)
 m_n^\tree(\sigma(1), \sigma(2),\ldots, \sigma(n)) \, ,\label{TreeAmp}
\end{equation}
where $S_n/Z_n$ is the set of all permutations, but with cyclic
rotations removed.  We have suppressed the dependence on the particle 
polarizations $\varepsilon_i$
and momenta $k_i$, but label each leg with the index $i$.  The
$T^{a_i}$ are fundamental representation matrices for the Yang-Mills
gauge group $SU(N_c)$, normalized so that ${\rm Tr}(T^aT^b) =
\delta^{ab}$. 
The behavior of color-ordered tree amplitudes as the momenta of
two color adjacent legs becomes collinear, is~\cite{mpReview}
\begin{equation}
m_n^{\tree}\ \mathop{\longrightarrow}^{a \parallel b}\
\sum_{\lambda=\pm}  
{\mathop{\rm Split}\nolimits}^{\tree}_{-\lambda}
(a^{\lambda_a},b^{\lambda_b})\,
      m_{n-1}^{\tree}(\ldots K^\lambda\ldots) \, ,
\label{TreeCollinear}
\end{equation}
where $\lambda$ represents the helicity, $m_n^{\tree}$ are color-decomposed
tree sub-amplitudes with a fixed ordering of legs and $a$ and
$b$ are consecutive in the ordering, with $k_a=zK$ and $k_b=(1-z)K$.
For the case of only gluons, the tree splitting functions
splitting into a positive helicity gluon
(with the convention that all particles are outgoing) is 
\begin{eqnarray}
{\mathop{\rm Split}\nolimits}^{\tree}_+(a^+,b^+)
             &=& 0 \,, \hskip 3.6 cm 
{\mathop{\rm Split}\nolimits}^{\tree}_+(a^-,b^-)
             = {-1\over \sqrt{z (1-z)} \spb{a}.{b} } \,, 
      \nonumber\\
{\mathop{\rm Split}\nolimits}^{\tree}_{+}(a^{-},b^{+})
             &=& {z^2\over \sqrt{z (1-z)} \spa{a}.{b} } \,, 
               \hskip 1. cm 
{\mathop{\rm Split}\nolimits}^{\tree}_{+}(a^{+},b^{-})
             = {(1-z)^2\over \sqrt{z (1-z)} \spa{a}.{b} } \,,
\label{TreeSplit}
\end{eqnarray}
where the remaining ones may be obtained by parity.  The spinor inner
products~\cite{SpinorHelicity,xu,mpReview} are $\spa{i}.j =
\langle i^- | j^+\rangle$ and $\spb{i}.j = \langle i^+| j^-\rangle$,
where $|i^{\pm}\rangle$ are massless Weyl spinors of momentum $k_i$,
labeled with the sign of the helicity.  They are antisymmetric, with
norm $|\spa{i}.j| = |\spb{i}.j| = \sqrt{s_{ij}}$, where $s_{ij} =
2k_i\cdot k_j$.

The behavior of color-ordered tree amplitudes in the soft limit is 
very similar to the
above.  As the momentum $k$ of an external leg becomes soft 
the color-ordered amplitudes become 
\begin{equation}
m_n^\tree (..., a,k^\pm, b,...)|_{k\to 0} =
{\rm Soft}^{\tree}(a,k^\pm, b)\, m_{n-1}^{\tree}(..., a, b,...)\,, 
\label{TreeSoft} 
\end{equation}
with the tree-level soft functions
\begin{equation}
{\rm Soft}^{\tree}(a,k^+,b) = {\left\langle a\,b\right\rangle
 \over \left\langle a\,k\right\rangle \left\langle k\,b\right\rangle}\,,
\hskip 2 cm 
{\rm Soft}^{\tree}(a,k^-,b) = {-[ a\,b]
\over [a\, k] [k\, b]}\,.
\end{equation}

The factorization of the collinear~(\ref{TreeCollinear}) and of the
soft~(\ref{TreeSoft}) limits are similar. However, 
due to the locality of the collinear emission, the factorization 
property~(\ref{TreeCollinear}) extends to the full amplitude~(\ref{TreeAmp}).
Conversely, because of the non-locality of the soft emission and of the
self-interactive nature of the gluon interaction, the 
factorization~(\ref{TreeSoft}) is true only at the color-ordered
amplitude level.

The color decomposition of one-loop multi-gluon
amplitudes with adjoint states circulating in the loop is
\cite{bkcolor}
\begin{equation}
M_n^\oneloop(1,2,\ldots n) =  g^n 
\sum_{j=1}^{\lfloor{n/2}\rfloor+1} \sum_{\sigma \in S_n/S_{n;j}}
 \Gr_{n;j}(\sigma) \, m_{n;j}^\oneloop(\sigma(1),\ldots,\sigma(n)) \,,
\label{OneLoopColor}
\end{equation}
where $\lfloor x\rfloor$ denotes the greatest integer less than or
equal to $x$, $\Gr_{n;1}(1) \equiv N_c \, \Tr\bigl(T^{a_1}\cdots
T^{a_n}\bigr)$, $\Gr_{n;j}(1) = \Tr\bigl(T^{a_1}\cdots
T^{a_{j-1}}\bigr) \, \Tr\bigl(T^{a_j}\cdots T^{a_n}\bigr)$ for $j>1$,
and $S_{n;j}$ is the subset of permutations $S_n$ that leaves the
trace structure $\Gr_{n;j}$ invariant, and where
$m_{n;j}^{\oneloop}$ are color-decomposed one-loop sub-amplitudes.
It turns out
that at one-loop the $m_{n;j>1}$ can be expressed in terms of
$m_{n;1}^\oneloop$~\cite{bdkReview}, so we need only discuss this case 
here.  The amplitudes with fundamental fermions in the loop 
contain only the $m_{n;1}^\oneloop$ color structures and are scaled by 
a relative factor of $1/N_c$.

The behavior of color-ordered one-loop amplitudes as the momenta of
two color adjacent legs becomes collinear, is~\cite{bddkSusy4,bdk3g2q}
\begin{equation}
m_{n;1}^{\oneloop}\ \mathop{\longrightarrow}^{a \parallel b}\
\sum_{\lambda=\pm}  \biggl\{
{\mathop{\rm Split}\nolimits}^{\tree}_{-\lambda}
(a^{\lambda_a},b^{\lambda_b})\,
      m_{n-1;1}^{\oneloop}(\ldots K^\lambda\ldots)
+{\mathop{\rm Split}\nolimits}_{-\lambda}^{\oneloop}
(a^{\lambda_a},b^{\lambda_b})\,
      m_{n-1}^{\tree}(\ldots K^\lambda\ldots) \biggr\} \, .
\label{OneLoopCollinear}
\end{equation}
The one-loop splitting functions are,
\begin{eqnarray}
\Split_+^\oneloop(a^-, b^-) &=& (\F+\N) \Split^\tree_+(a^-, b^-)\,, \nonumber\\
\Split_+^\oneloop(a^\pm, b^\mp) &=& \N\,  \Split^\tree_+(a^\pm, b^\mp)\,,
\label{SplitLoop} \\
\Split_+^\oneloop(a^+,b^+) &=& -\F \,{1\over \sqrt{z (1-z)}}  
 {\spb{a}.{b}\over \spa{a}.b^2}\,. \nonumber
\end{eqnarray}
The function $\F$ arises from the `factorizing' contributions
and the function $\N$ arises from the `non-factorizing' ones 
described in ref.~\cite{bc} and are given through $\Ord(\eps^0)$
by~\cite{bddkSusy4,bdk3g2q}
\begin{eqnarray}
\F &=& {1\over 48 \pi^2} 
 \Bigl(1 - {N_{\! f}\over N_c} \Bigr) z (1-z) + \Ord(\eps)\, ,
\label{FNFuncs}\\
\N &=& 
c_\Gamma \Bigl[
- {1 \over \epsilon^2} 
\Bigl( {\mu^2 \over z(1-z) (-s_{ab})}\Bigr)^\epsilon
 + 2 \ln (z) \ln(1-z)
  - {\pi^2 \over 6}\Bigr] + {\cal O}(\epsilon) \, ,\nonumber
\end{eqnarray}
with $N_{\! f}$ the number of quark flavors and
\begin{equation}
c_{\Gamma} = {1\over 
(4\pi)^{2-\epsilon}}\, {\Gamma(1+\epsilon)\,
\Gamma^2(1-\epsilon)\over \Gamma(1-2\epsilon)}\, .\label{cgam}
\end{equation}
As at tree-level, the remaining splitting functions can be obtained 
by parity.  The
explicit values were obtained by taking the limit of five-point
amplitudes; the universality of these functions for an arbitrary
number of legs was proven in ref.~\cite{bc}.

The functions (\ref{FNFuncs}) 
have been extended to all orders in $\eps$ in ref.~\cite{bds}
\begin{eqnarray}
\F &=& {2 \cg \over (3-2\eps) (2-2\eps) (1-2\eps)} 
\Bigl(1 -\eps\delta_R - {N_{\! f}\over N_c} \Bigr) \, 
\Bigl( {\mu^2 \over - s_{ab}} \Bigr)^\eps  z(1-z) \,, 
\label{FNFuncsAllEps} \\
\N &=&  c_{\Gamma}\, 
\left({\mu^2\over -s_{ab}}\right)^{\epsilon}\, {1\over\epsilon^2} 
\left[ -\left({1-z\over z}\right)^\epsilon 
{\pi\epsilon\over \sin(\pi\epsilon)}\,
+ 2 \sum_{k=1,3,5,...} \epsilon^k \,
{\rm Li}_k\left({-z\over 1-z}\right) \right]\, , \nonumber
\end{eqnarray}
where the polylogarithms are defined as \cite{lewin}
\begin{equation}
\left. \begin{array}{l}
{\rm Li}_1(z)\ = - \ln(1-z) \\ \displaystyle
{\rm Li}_k(z) = \int_0^z {dt\over t}\,{\rm Li}_{k-1}(t) \qquad (k=2,3,\dots)
\end{array}\right\} = \sum_{n=1}^{\infty} {z^n\over n^k}\, ,
\label{polys} 
\end{equation}
and the regularization scheme  parameter is,
\begin{equation}
\delta_R = \left\{ \begin{array}{ll} 1 & \mbox{HV or CDR scheme},\\
0 & \mbox{FDH or DR scheme}, \end{array}
\right. \, \label{cp}
\end{equation}
where CDR denotes the conventional dimensional regularization scheme,
HV the 't Hooft-Veltman scheme, DR the dimensional reduction scheme,
and FDH the `four-dimensional helicity scheme.  (For further
discussions on scheme choices see refs.~\cite{bk,cst}.) 

The behavior of one-loop amplitudes in the soft limit, as 
the momentum $k$ of an external leg becomes soft, is given by
\begin{eqnarray}
\lefteqn{ m_{n;1}^\oneloop (..., a,k^\pm, b,...)|_{k\to 0} =} 
\label{OneLoopSoft}\\ 
& & {\rm Soft}^{\tree}(a,k^\pm, b)\, m_{n-1;1}^\oneloop(..., a, b,...) 
+ {\rm Soft}^\oneloop (a,k^\pm, b)\, m_{n-1}^{\tree}(..., a, b,...)\,, 
\nonumber
\end{eqnarray}
where the one-loop gluon soft function may be
extracted through $\Ord(\epsilon^0)$ from four- \cite{bk} and five-parton
\cite{bdk5g,kst,bdk3g2q} one-loop amplitudes, and it is
\begin{equation}
{\rm Soft}^{\oneloop}(a,k^\pm,b) = - {\rm Soft}^{\tree}(a,k^\pm,b)\,
c_{\Gamma}\, \left({\mu^2(-s_{ab})\over (-s_{ak})(-s_{kb})}
\right)^{\epsilon}\, \left({1\over\epsilon^2} + {\pi^2\over 6} \right)
+ {\cal O}(\epsilon)\, .\label{softexp}
\end{equation}
Eq.~(\ref{softexp}) does not depend on $N_{\! f}$ or $\delta_R$.
In ref.~\cite{bds} we have extended it to all orders of $\epsilon$, 
with the result, 
\begin{equation}
{\rm Soft}^{\oneloop}(a,k^\pm,b) = - {\rm Soft}^{\tree}(a,k^\pm,b)\,
c_{\Gamma}\, {1\over\epsilon^2}\,
\left({\mu^2(-s_{ab})\over (-s_{ak})(-s_{kb})}\right)^{\epsilon}\, 
{\pi\epsilon\over \sin(\pi\epsilon)}\, .\label{soft}
\end{equation}

We now apply the results for the
soft function (\ref{soft}) to the case of three-gluon production 
in multi-Regge kinematics.  To do so, 
we also need the four-gluon one-loop amplitude through 
$\Ord(\eps)$. In fact, this is known exactly to all orders of $\epsilon$. 
In the high-energy limit, $s\gg t$, its dispersive part, which is all
that contributes to the NLL BFKL kernel, is~\cite{bds}
\begin{eqnarray}
\lefteqn{ {\rm Disp}\, M_4^\oneloop(A^-,A'^+,B'^+,B^-) =  
M_4^{\tree}(A^-,A'^+,B'^+,B^-)\, g^2\, c_{\Gamma}\, 
\left({\mu^2\over - t}\right)^{\epsilon}\,
{1\over \epsilon (1-2\epsilon)} } \label{alleps}\\ &\times&
\Biggl\{ N_c\, \left[2(1-2\epsilon) \left(\psi(1+\epsilon) - 
2\psi(-\epsilon) + \psi(1) + \ln{s\over -t}\right) + {1 - 
\delta_R\epsilon\over 3-2\epsilon}
- 4\right] + {2(1-\epsilon)\over 3-2\epsilon} N_{\! f} \Biggr\}\, , 
\nonumber
\end{eqnarray}
where $A,\,B$ and $A',\,B'$ are respectively the incoming and outgoing
gluons. The unrenormalized five-gluon one-loop 
amplitude in the multi-Regge kinematics, and in the soft
limit for the intermediate gluon and to all orders in $\epsilon$, is 
obtained by using eq.~(\ref{OneLoopSoft}), with the four-gluon one-loop 
amplitude~(\ref{alleps}), and the dispersive part of the soft 
function~(\ref{soft}), yielding~\cite{bds}
\begin{eqnarray}
\lefteqn{ {\rm Disp}\, M_5^\oneloop(A^-,A'^+,k^\pm,B'^+,B^-)\bigr|_{k\to 0} =  
 g^2\, c_{\Gamma}\, M_5^{\tree}(A^-,A'^+,k^\pm,B'^+,B^-)\bigr|_{k\to 0}\, } 
\nonumber\\ &\times& \Biggl[ \left({\mu^2\over - t}\right)^{\epsilon}
\Biggl\{ N_c\, \Biggl[-{4\over\epsilon^2} + {2\over\epsilon}\,
\left(\psi(1+\epsilon) - 2\psi(1-\epsilon) + \psi(1) + \ln{s\over -t}\right) 
\label{softlim}\\ &+& {1\over \epsilon (1-2\epsilon)}\,\left({1 - 
\delta_R\epsilon\over 3-2\epsilon} - 4\right)\Biggr] + 
{2(1-\epsilon)\over \epsilon (1-2\epsilon)(3-2\epsilon)} N_{\! f} \Biggr\} 
\nonumber\\ &-& N_c \left({\mu^2\over 
|k_{\perp}|^2}\right)^{\epsilon}\, {1\over\epsilon^2}\,
\left[1 + \epsilon \psi(1-\epsilon) - \epsilon \psi(1+\epsilon)\right]
\Biggr]\, ,\nonumber
\end{eqnarray}
which agrees through $\Ord(\epsilon^0)$ with 
the five-gluon one-loop amplitude, with strong rapidity ordering and 
in the soft limit for the intermediate gluon \cite{bdk5g,dds}.
Eq.~(\ref{softlim}) can than be matched to the full five-gluon one-loop 
amplitude, with strong rapidity ordering, computed through $\Ord(\epsilon^0)$.
The result~\cite{dds} agrees with the NLL corrections to the Lipatov vertex
computed in ref.~\cite{ffk} in the CDR scheme, through $\Ord(\epsilon)$.

In conclusion, in this talk we have examined how one-loop soft and collinear 
splitting functions occur in the calculation of NNLO corrections to
production rates, and we have presented the one-loop gluon soft and 
splitting functions, computed to all orders in $\epsilon$. 
We have then applied the one-loop gluon soft function to the calculation
of the NLL corrections to the Lipatov vertex to all orders in 
$\epsilon$\cite{bds}.
A systematic discussion of the soft and collinear splitting
functions, including the case of external fermions, is
presented elsewhere~\cite{bdks}.

\vskip .2 cm 

This work was supported by the US Department of Energy under grants
DE-FG03-91ER40662 and DE-AC02-98CH10886, by the US National Science 
Foundation under grant
PHY-9722144 and by the EU Fourth Framework Programme {\em Training and
Mobility of Researchers}, Network {\em Quantum Chromodynamics and the 
Deep Structure of Elementary Particles}, contract FMRX-CT98-0194 
(DG 12 - MIHT). The work of V.D.D. and C.R.S was also supported by
NATO Collaborative Research Grant CRG-950176.


\end{document}